\documentclass[aps,prl,twocolumn,nofootinbib,epsfig]{revtex4}
\usepackage{amsmath,amssymb,latexsym}
\usepackage{graphicx}
\usepackage{bm}
\def\be{\begin{equation}}
\def\ee{\end{equation}}
\def\ed{\end{document}}

\begin{document}

\title{Footprints of Super-GZK Cosmic Rays in the Pilliga State Forest}
\author{Luis Anchordoqui and Haim Goldberg}
\affiliation{Department of Physics, Northeastern University, Boston, MA 02115}

\begin{abstract}
High energy cosmic ray data collected by the SUGAR Collaboration in the Pilliga
State Forest  have been re-analyzed using up-to-date shower simulations.
Complete, inclined angle, Monte Carlo simulations reveal 2 events with 
energies in excess of $7\times 10^{19}$ eV at 95\%CL, independent of the 
choice of hadronic interaction
model and of chemical composition of the primary. An additional 3 events, with 
mean
energy $\agt 7\times 10^{19}$ eV, were also re-analyzed in the same manner.
A lower bound on the flux at the high end of the spectrum, as observed
in the Southern sky, is presented on the basis of our analysis.

\centerline{NUB-3241-TH-03}

\end{abstract}

\maketitle

\section{Introduction}
In the mid-60's Greisen, Zatsepin, and Kuzmin (GZK) pointed out that
if cosmic rays originate at cosmological distances,
standard physics implies there would be an ultra violet cutoff in the
observed spectrum~\cite{Greisen:1966jv}.  Due to a lack of
knowledge of the primary species, nowadays there is some ambiguity in the
definition of the GZK energy limit.  Gamma-rays produce
electromagnetic cascades and for extragalactic magnetic fields ${\cal O}$(nG)
the high energy component of the spectrum is dramatically depleted.
The net effect of proton interactions would be a pile-up around
$5 \times  10^{19}$~eV
with the spectrum dropping sharply thereafter. Nucleus photodisintegration is
particularly important at energies which excite the giant dipole resonance,
yielding a
cutoff which is species-dependent and is slightly shifted to higher
energies~\cite{Bhattacharjee:1998qc}. With this in mind, in this work
we consider an event to supersede the cutoff if its lower energy limit at the
95\%~CL exceeds $7 \times 10^{19}$~eV. This conforms closely to the most
conservative criteria delineated in Ref.~\cite{Farrar:1999fw}.

A first hint of a puzzle surfaced in the  highest energy Fly's Eye
event~ \cite{Bird:1994uy} which has no apparent progenitor
within the local supercluster~ \cite{Elbert:1994zv}. Subsequent observations
with the Akeno Giant Air Shower Array (AGASA)~\cite{Takeda:1998ps}
carried strong indication that
the cutoff was somehow circumvented in the absence of plausible nearby
sources. Far from adding confirmation to what seemed
a fascinating discovery, the recent analysis~\cite{Abu-Zayyad:2002sf} of
data recorded in monocular
mode by the HiRes experiment has re-infused the field with
uncertainty.

An analysis~\cite{Bahcall:2002wi} of the combined data reported by the HiRes,
the Fly's Eye, and the Yakutsk collaborations is supportive of the existence
of the GZK cutoff at the  $>5\sigma$  ($>3.7\sigma$) level. The deviation
from GZK depends on the set of data used as a basis for power law
extrapolation from lower energies. In additional input to this analysis,
it has been recently noted~\cite{Watson} that there may be technical problems
with the Yakutsk data collection. We have nothing further to add to this
discussion, but it seems worthwhile to analyze any other independent set of
data.

\section{SUGAR data revisited}
The Sydney University Giant Air-shower Recorder (SUGAR) is the largest
array ever built in the Southern hemisphere. It ceased operation in February
1979 after a life of 11 years~\cite{Winn:un}. The experiment comprised 47
detector stations
located in the Pilliga State Forest (New South Wales, Australia) at latitude
$30^\circ$ 31' S, longitude $149^\circ$ 38' E, and altitude 250 m above sea
level. Each of these stations consisted of two 6~m$^2$ conical liquid
scintillator tanks separated by 50~m and buried under $1.5 \pm 0.3$~m of
earth to shield against the electromagnetic component of the
shower~\cite{Brownlee}.
A minimum ionizing muon traversing the scintillator produced an average of
10 photoelectrons, which were detected by photomultiplier tubes (PMTs)
looking downward at the scintillator. Each tank was equipped with
discriminators whose
thresholds were set to the signal size expected for 3 minimum ionizing muons.
A local trigger was generated for the case of hits on both tanks within 350~ns
of one another. In this work we are interest in the largest showers, which
determine the high end of the spectrum (energy $\agt 7 \times 10^{19}$~eV).
In order to obtain a clean sample we
only analyze events with the shower zenith angle $< 45^\circ$ and
muon sizes $> 10^8$ particles. Up to 1000~m from the core all stations in
these showers gave non-zero readings  and thus well determined densities can
be given. We require a global trigger of 6 or more stations which did record
local events,  with a minimum of 8 stations triggered within a time window
of $80.5~\mu$s. There are 2 events (serial \#12420 and \#6179) which
survive all these cuts.

A first estimate of the shower arrival direction is made by fitting
a planar shower front to the station hit times.  A core search is then
carried out using an iterative procedure.  Next, the lateral distribution
function is fitted to the observed signals in the stations. In fitting the
muon density account is taken of both stations which did record local
events, and those which did not. The SUGAR analysis fitted using
a Fisher~\cite{Fisher} function, which  is a modified version of the
Greisen~\cite{Bennett}
muon structure function with the exponent of the distant term being zenith
angle dependent,
\begin{equation}
\rho_\mu(r) = {\cal N}\,k(\theta)\,
\left(\frac{r}{r_0}\right)^{-a}\,\left( 1 + \frac{r}{r_0}\right)^{-b}\,\,.
\end{equation}
Here, ${\cal N}$ is a normalization constant,  $\theta$
is the incident zenith angle, $r$ is the perpendicular
distance from the shower axis, $r_0 = 320$~m, $a = 0.75$,
$b= 1.50 + 1.86 \cos\theta$, and
\begin{equation}
k(\theta) =  \frac{1}{2\pi r_0^2}\,
\frac{\Gamma(b)}{\Gamma(2-a)\,\Gamma(a+b-2)}\,.
\end{equation}
In our  analysis, we extract the lateral distribution directly from
Monte Carlo simulations of showers
using {\sc aires} 2.6.0~\cite{Sciutto:1999rr}.
For all shower simulations, we specify the geographical coordinates of SUGAR to
turn on the geomagnetic field library in {\sc aires}~\cite{geo}. Moreover,
we discard muons with energies below the muon detection threshold of a SUGAR
scintillator, or $(0.75 \pm 0.15)\sec \theta$~GeV.

A first source of systematic uncertainties is encoded in the hadronic
interaction model used to process the high energy
collisions. In the eikonal model~\cite{Glauber:1970jm} of high energy
hadron-hadron scattering, the unitarized cross section is written as
\begin{equation}
\sigma_{inel}=\int d^2\vec b\,
\left(1-\exp\left\{ -2\chi_{_{\rm soft}}(s,\vec b)
-2\chi_{_{\rm hard}}(s,\vec b)\right\}\right)\ ,
\end{equation}
where the scattering is compounded as a sum of ladders via hard
and soft processes through the eikonals $\chi_{_{\rm hard}}$ and
$\chi_{_{\rm soft}}$. The leading contenders to approximate the
(unknown) cross sections at cosmic ray energies, {\sc
sibyll}~\cite{Fletcher:1994bd} and {\sc
qgs}jet~\cite{Kalmykov:te}, share the eikonal approximation but
differ in their {\em ans\"atse} for the eikonals. In both cases,
the core of dominant scattering at very high energies is the
parton-parton minijet cross section. In the {\sc qgs}jet model,
this core of the hard eikonal is dressed with a soft-pomeron
preevolution factor; this amounts to take a parton distribution
which is Gaussian in the transverse coordinate distance $|\vec
b|$. In  {\sc sibyll}, the transverse density distribution is
taken as the Fourier transform of the electric form factor,
resulting in an exponential (rather than Gaussian) falloff of the
parton density profile with $|\vec b|$. Some of the shower
characteristics resulting from these choices are readily
predictable: the harder form of the {\sc sibyll} form factor
allows a greater retention of energy by the leading particle, and
hence less available for the ensuing
shower~\cite{Anchordoqui:2000uh}. This in turn explains the
reduction of mean secondary multiplicity
observed~\cite{Anchordoqui:1998nq} in going from {\sc qgs}jet to
{\sc sibyll} simulations. The growth in inelastic cross section
saturates the $\log^2s$ Froissart bound in both cases, but the
multiplicative constant turns out to be larger for {\sc sibyll},
enhancing the cross section by about 30\% at energies of
$10^{19-20}$~eV.  However, the shower characteristics are the
result of exponentially many hadronic collisions in air, and for
these the larger {\sc sibyll} cross section is somewhat offset by
its smaller multiplicity, reducing the differences between the two
schemes down to the level of 10-20\%. These and other properties
of the two schemes have been amply displayed in a recent
work~\cite{Alvarez-Muniz:2002ne}.

\begin{figure}
\begin{center}
\includegraphics[height=8.cm]{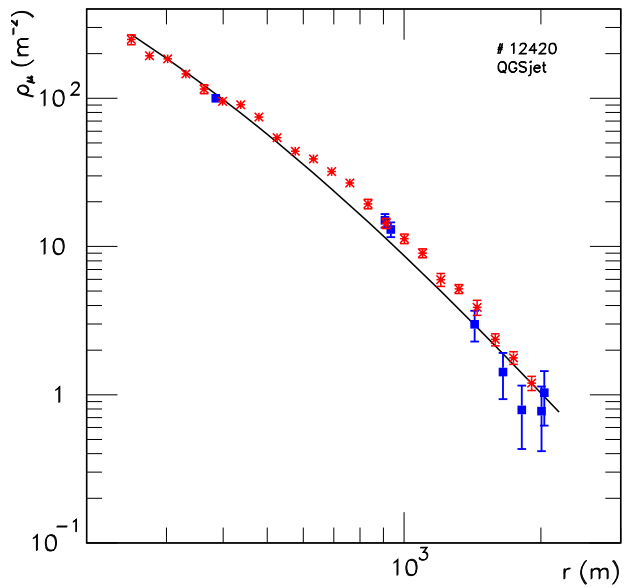}
\caption{Best fit to the muon density  ($*$)
for event \#12420 using {\sc aires + qgs}jet.
The estimated primary energy is
$1.1 \times 10^{20}$~eV,  with $\chi^2/{\rm d.o.f.} = 9.4/7$.  The recorded particle densities as
reported by the SUGAR Collaboration are indicated with squares. The solid line
indicates the best fit reported by SUGAR Collaboration using the Fisher
function ($\chi^2/{\rm d.o.f.} = 10.3/7$).} \label{fig12420qgsjet}
\end{center}
\end{figure}

\begin{figure}
\begin{center}
\includegraphics[height=8.cm]{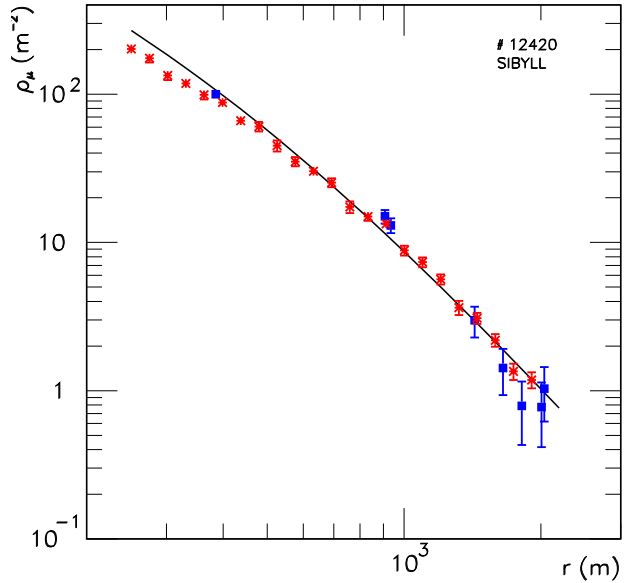}
\caption{Best fit to the muon density  ($*$)
for event \#12420 using {\sc aires + sibyll}.
The estimated energy is \mbox{$1.2 \times 10^{20}$~eV,}
with \mbox{$\chi^2/{\rm d.o.f.} = 9.1/7$}. The recorded particle densities as
reported by the SUGAR Collaboration are indicated with squares. The solid line
indicates the best fit reported by SUGAR Collaboration using the Fisher
function \mbox{($\chi^2/{\rm d.o.f.} = 10.3/7$)}.} \label{fig12420sibyll}
\end{center}
\end{figure}

Eight stations were involved in the event \#12420, all with local
triggers (i.e., 16 scintillators).  A map of the recorded muon
densities in a plane perpendicular to the shower axis was
reported by the SUGAR Collaboration in Ref.~\cite{Bell}. The
north tank of one of the stations recorded a saturated response
(density $ > 800$~m$^{-2}$). The standard analysis of the SUGAR
Collaboration placed the shower core on this tank. However, the
collaboration concluded that in such a big shower saturated
densities may occur at distances up to several hundred meters
from the core. Moreover, for large signals, afterpulsing in the
PMTs can lead to an overestimate of the muon density. The
location of the core was best re-estimated by assigning lower
energy densities to this station. Since the probability of
recording a saturated response from an actual density less than
100~m$^{-2}$ is negligible, the SUGAR Collaboration concluded
that this value of density in both tanks would yield a minimum
value on the primary energy. In our analysis we adopt this
conservative lower bound. Note that this second source of
systematic error can lead to an underestimate of the energy. The
best $\chi^2$ fit for {\sc qgs}jet01 ({\sc sibyll} 2.1), assuming
a proton primary, is given in Fig.~\ref{fig12420qgsjet}
(Fig.~\ref{fig12420sibyll}). Note  that {\sc aires + sibyll}
simulations result in an energy which exceeded the {\sc aires +
qgs}jet value by about 10\%. This agrees with the analysis of the
AGASA Collaboration~\cite{Takeda:2002at}.

\begin{figure}
\begin{center}
\includegraphics[height=8.cm]{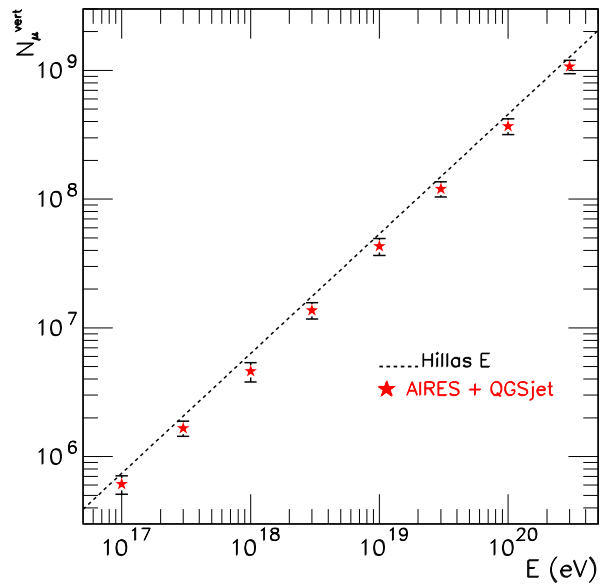}
\caption{Total number of muons as a function of the energy shower. The
simulations were carried out with {\sc aires + qgs}jet, assuming protons
as primaries. The particles were injected vertically at the top of the
atmosphere and the detector was placed at 250~m above sea level. Muons
with energies below the SUGAR threshold  (0.75~GeV) are not taken into
account in the simulations. The dashed line
indicates the Hillas parametrization (see main text).}
\label{hillas}
\end{center}
\end{figure}

The energy that we find for event\#12420 using {\sc qgs}jet is about 5\%
above the energy reported by the SUGAR Collaboration~\cite{Winn:un},
their estimate being based on the Hillas E model~\cite{Hillas}
\begin{equation}
E_{\rm SUGAR} =
1.64\times 10^{18} \,
\left(\frac{N_{\mu}^{\rm vert}}{10^7}\right)^{1.075}\, \, {\rm eV} \, .
\label{hs}
\end{equation}
In further exploring the energy systematics, we have generated
proton vertical showers using {\sc aires+qgs}jet and compared
(in Fig.~\ref{hillas}) the $N_{\mu}^{\rm vert}-E$ relation with that given in
Eq.~(\ref{hs}). As can be
seen from Fig.~\ref{hillas}, this conversion formula underestimates the
energy by about 25\%, for a given $N_{\mu}^{\rm vert}$, when compared with
our simulations. However, as noted above, the overall
underestimate of the SUGAR assignment for the energy from our
best fit is about 5\%. This is readily explained by noting that
the normalization constant ${\cal N}$ (used by the SUGAR
Collaboration as the muon size) and obtained through the fit to
the Fisher function overestimates the number of muons (as found
through our simulation) by about 17\% (see Table~\ref{tt}). Since the
equivalent
vertical muon size and energy bear a near-linear relationship to
one another (see Fig.~\ref{hillas}), these systematic effects nearly offset
one another, bringing the SUGAR energy estimate near to that
obtained with {\sc aires+qgs}jet. Implicitly, this discussion supports
the consistency of the constant intensity cut method of mapping
muon sizes from inclined to vertical~\cite{Alvarez-Muniz:2002xs}.

A third source of systematic uncertainty is hidden in the
chemical composition. As a first approximation, a shower produced
by a nucleus with energy $E_{_A}$ and mass $A$ can be modeled by
the collection of $A$ proton showers, each with $A^{-1}$ of the
nucleus energy. Now, using Eq.~(\ref{hs}) a straightforward
calculation shows that the total number of muons produced by the
superposition of $A$ individual proton showers is, $N_{\mu}^{\rm
vert} \propto A (E_{_A}/A)^{0.93}$. Consequently, one expect a
cosmic ray nucleus to produce about $A^{0.07}$  more muons than a
proton. Conversely, a given muon size will imply a smaller energy
for a complex nucleus. Thus, we can  estimate the lower energy
limit of the event \#12420, by simulating several sets of showers
for $^{56}$Fe  primaries~\cite{iron}. The {\sc qgs}jet package
was used to process the high energy hadronic interactions.
Energies below $7.0 \times 10^{19}$~eV are excluded at the
95\%~CL.

The analysis of the second  event (\#6179) is less
straightforward. This is because there were stations in which one
of the tanks did not fire~\cite{Bell}. In assigning the trigger
intensity of
these tanks, the SUGAR Collaboration took as a maximum the
density in the firing tanks, and as  a minimum the density
averaged over all the tanks that were expected to fire. Data scaled
to a muon size of $4.5\times 10^8$ is shown in Ref.~\cite{Bell}. Because
of the resulting decrease in the quality of the statistics, the
likelihood in this case did not converge reliably to a global
minimum. However, a 95\% confidence interval, $7.0\times
10^{19}~{\rm eV} < E < 1.2\times 10^{20}~{\rm eV},$ was readily
obtained.

\begin{figure}
\begin{center}
\includegraphics[height=8.5cm]{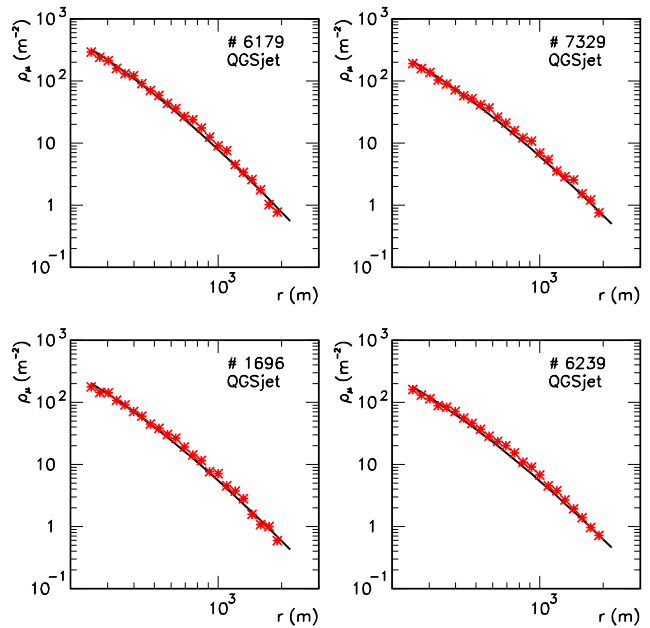}
\caption{Lateral distributions obtained using air-shower simulations
with {\sc aires + qgs}jet ($*$) superimposed over the best fits of the
Fisher function (solid line) as reported by the SUGAR Collaboration. The
corresponding primary
energies used in the simulations are those given in Table~\ref{tt}.}
\label{sugarqgsjet}
\end{center}
\end{figure}

For a more extended comparison with the analysis reported by the
SUGAR Collaboration, we relax the global trigger designated
above, and follow  several more recent analyses of SUGAR
data~~\cite{Kewley:zt} by considering events in which 5 pairs of
scintillators fired within a time window of 80.5 $\mu$s. For this
kind  of global trigger, the uncertainty in the arrival direction
is manageable at an average of 4.3$^{\circ}$~\cite{Brownleephd}.
(For 3- and 4-fold triggers, this rises beyond 10 degrees.) With
this new global trigger, we simulate showers with {\sc aires +
qgs}jet and find that by increasing the energy reported by the
SUGAR Collaboration by about 5\% we reasonably conform to the
lateral profiles of the events as described by the Fisher
function, with normalization constants taken from~\cite{Winn:un}
and given in Table~\ref{tt}. As a result, we find 4  events (including
\#6179 discussed above) with energies $> 7\times 10^{19}$~eV.
These qualitative fits are shown in  Fig.~\ref{sugarqgsjet}, and the parameters
characterizing these events are detailed in Table~\ref{tt}.

\begin{table}
\caption{Primary energy in units of $10^{20}$~eV for proton primaries
estimated using {\sc aires + qgs}jet. The corresponding muon sizes and
normalization constant of the Fisher function are also given in units
of $10^8$ particles. The last 3 columns
indicate the incident zenith angle and the arrival directions in
equatorial coordinates ($\alpha,\delta$) with respect to the
reference system B1950~\cite{Winn:un}. The estimated angular uncertainty
for all these events is $4.3^\circ$.}
\begin{tabular}{|c|c|c|c|c|c|c|}
\hline
\hline
Number &$\,\,\,\,\,{\cal N}\,\,\,\,\,$ &
$\,\,\,\,\,E\,\,\,\,\,$ & $\,\,\,\,\,N_\mu\,\,\,\,\,$ &  $\theta$ [deg.]
& $\alpha$ [deg.] & $\delta$ [deg.]  \\
\hline
\hline
12420 & 3.47 & 1.10 & $2.99 \pm 0.35$   & 43  &  147.0 & $-43.6$ \\
6179 & 4.06  & 1.03 & $3.56 \pm 0.45$  &  24 &  121.5 &  $-32.0$ \\
7329 & 2.63  & 0.77 & $2.24 \pm 0.34$  &  38 &  135.0 &  $\phantom{-}04.3$ \\
1696 & 2.55  & 0.71 & $2.20 \pm 0.27$  &  33 &  331.5 &  $-32.2$ \\
6239 & 2.27  & 0.70 & $1.96 \pm 0.27$  &  40 &  280.5 & $-55.6$  \\
\hline
\hline
\end{tabular}
\label{tt}
\end{table}

\begin{figure}
\begin{center}
\includegraphics[height=8.5cm]{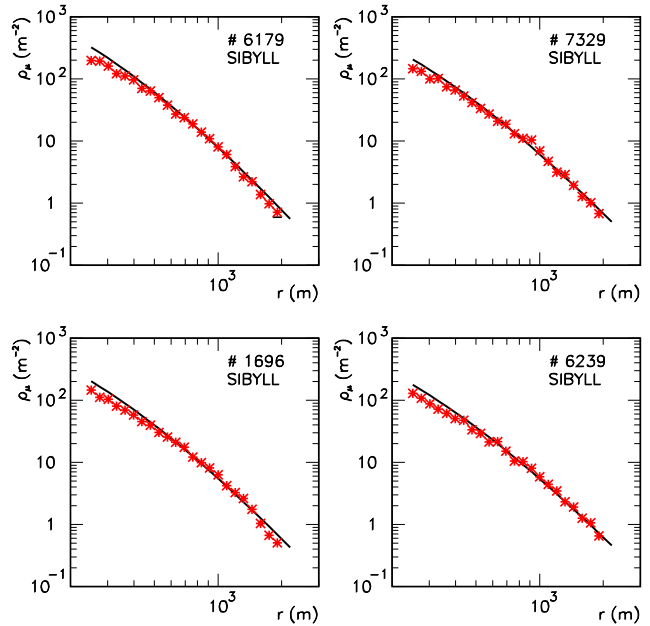}
\caption{Lateral distributions obtained using air-shower simulations
with {\sc aires + sibyll} ($*$) superimposed over the best fits of the
Fisher function (solid line) as
reported by the SUGAR Collaboration. The corresponding primary energies in
units of $10^{19}$~eV used in the simulations are: 11.3, 8.5, 7.8, and 7.7,
for \#6179, \#7329, \#1696 and \#6239, respectively.} \label{sugarsibyll}
\end{center}
\end{figure}

A closer  examination of Fig.~\ref{sugarqgsjet}  reveals a systematic
deviation between the {\sc qgs}jet and Fisher function lateral
distribution profiles. As can be seen in Fig.~\ref{sugarsibyll}, the
divergences become more severe when {\sc sibyll} is used to process the
hadronic interactions. Of course, this comparison with the Fisher
function plays no role in discriminating between these schemes.

\section{Energy spectrum}
The effective detecting area of the array at any time is given by the product
of the active geometric
surface area of the array, $S$, times the
probability $p({\cal N}, \theta)$ that a shower falling within $S$ will
be detected,
\begin{equation}
A_{\rm eff}({\cal N}, \theta, t) = S(t)\,\, p({\cal N}, \theta) \,.
\end{equation}
The SUGAR Collaboration reported~\cite{Horton} an angle specific exposure
for showers with ${\cal N} = 4 \times 10^8$ and $\theta = 32^\circ$,
\begin{equation}
{\cal E} (32^\circ) = \int_0^{T} \,dt\, S(t)\, p(4 \times 10^8, 32^\circ) =
540 \,\,{\rm km}^2 {\rm yr}\,,
\label{exposure}
\end{equation}
where $T = 11$ yr. This value of $p$ was obtained by allowing a global trigger of
3 or more stations.
Because the detection probability  decreases with  smaller
muon size (as is the case for the events in Table~\ref{tt}) and with our restriction to 5 triggers,
Eq.~(\ref{exposure}) leads to an upper limit on the integrated exposure for the
parameter space relevant to this study,
\begin{eqnarray}
{\cal E} (\theta < 45^\circ) & = & 2 \pi \,\,{\cal E} (32^\circ)\, \int_{\sqrt{2}/2}^1
\frac{p(\theta)}{p(32^\circ)}\, \cos\theta \,\, d \cos \theta \nonumber\\
 & \approx &
862~{\rm km}^2\,\,{\rm sr\,\, yr},
\end{eqnarray}
where $p(\theta) \propto \cos \theta$ is the probability of detection at
zenith angle $\theta$~\cite{Brownlee2}.

The observed number of events in an energy bin $\Delta$ is given by
\begin{eqnarray}
N &= &{\cal E} \int_{\Delta} E J(E) \frac{dE}{E}  \nonumber \\
&\simeq& 2.3 \,{\cal E}\, \,  \overline{E} J(\overline{E})\, \,\,\Delta\log_{10}E\ \ ,
\label{count}
\end{eqnarray}
so that
\begin{equation}
 \overline{E}^3 J(\overline{E}) = \overline{E}^2\frac{N}{2.3\,\,{\cal E}\,\,\,
\Delta\log_{10}E}\ \ ,
\label{flux}
\end{equation}
where $\overline{E}$ is the value of $E$ at the center of the logarithmic interval.

\begin{figure}
\begin{center}
\includegraphics[height=8.cm]{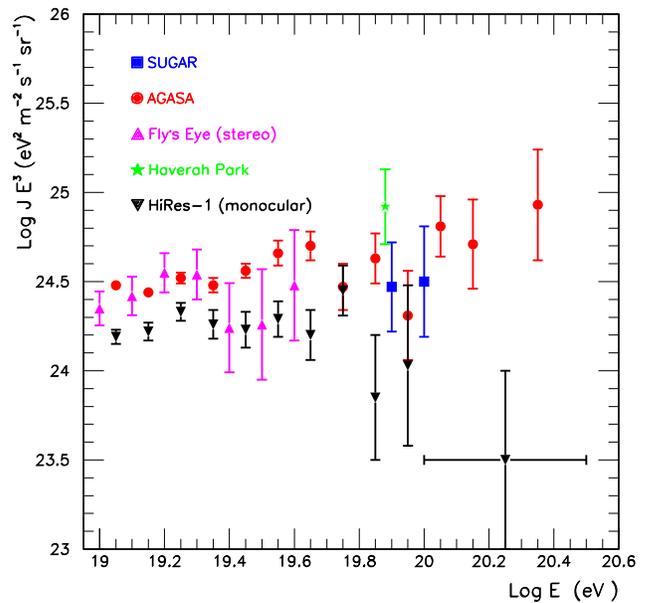}
\caption{Upper end of the cosmic ray energy spectrum as observed by 5
different experiments (for details see main text).}
\label{azukar}
\end{center}
\end{figure}

Following common practice, we adopt a bin size $\Delta\log_{10}E =0.1.$
As previously discussed, we have adopted a lower energy cut of $10^{19.85}$ eV.
The events listed
in Table~\ref{tt} are then grouped in two bins, and the resulting
contributions to the cosmic ray spectrum as given
by Eq.~(\ref{flux}) are shown in Fig.~\ref{azukar}. The error on the mean
\begin{eqnarray}
\Delta \log_{10}[\overline{E}^3 J(\overline{E})]  & = & \Delta\log_{10} N \nonumber\\
 & = & 0.343 \,\, \Delta \ln N \nonumber \\
 & = & 0.343\,\, \frac{\Delta N}{N} \,,
\end{eqnarray}
is taken as Poisson (i.e., $\Delta N = \sqrt N$).

On the same figure are shown the data points for the flux as reported
by the AGASA~\cite{Takeda:2002at}, Fly's Eye~\cite{Bird:wp}, and
HiRes~\cite{Abu-Zayyad:2002sf} collaborations. Also shown is the flux obtained
by a recalibration of the 4 highest energy events observed at the Haverah Park
facility~\cite{Ave:2001hq}. We note that the AGASA flux plotted is taken
from the
most recent analysis published by the AGASA
Collaboration~\cite{Takeda:2002at}, containing events
with zenith angle $< 45^{\circ},$ recorded until July 2002. The energy is
calibrated
using an average over hadronic interaction models and primary species
($p$ and $^{56}$Fe). It is of interest to the present work that the energies
obtained in this manner differ by only about 2\% with respect to those
obtained via {\sc aires + qgs}jet simulation.

\section{Summary and discussion}
\noindent $(1)\quad$ We have re-examined cosmic ray data taken over two
decades ago at the SUGAR facility, using up-to-date simulation methods to
describe shower evolution in the atmosphere, with resulting correlation
between muon size and energy. We conclude that there is strong evidence for
the existence of events superseding the GZK cutoff. Specifically, there are
two events in the sample with energy $> 7\times 10^{19}$ eV at the 95\%CL.
This statement includes consideration of systematic uncertainties
stemming from choice of hadronic interaction model and chemical composition.

\noindent $(2)\quad$ As part of our exploration of the systematics, we have
compared our study with methods used by the SUGAR Collaboration for energy
assignment. We found that through a compensation of errors in assessing muon
size from fits to the Fisher function and the normalization of the Hillas
relation~\cite{Hillas}, the final SUGAR energies are consistently about
5\% below those  found via {\sc aires + qgs}jet simulation. Remarkably, we
find (Fig.~\ref{hillas}) that the exponent in the power law
$N_{\mu}^{\rm vert}= {\cal K} E^{\alpha}$ is consistent with
Hillas's $\alpha = 0.93$, but with a shift in
the constant ${\cal K}$ upward by 25\%. A detailed recent
study~\cite{Alvarez-Muniz:2002ne} shows that in fact the exponent
$\alpha$ is more properly taken to have a logarithmic energy-dependence.

\noindent$(3)\quad$ Using a conservative estimate on the acceptance of the
SUGAR facility, we have been able to present lower bounds on the cosmic ray
flux in the case of a proton primary with  {\sc qgs}jet as the hadronic
interaction model. This lower bound would prevail for {\sc sibyll} independent
of the primary composition. The lower bounds would be slightly modified in the
case that {\sc qgs}jet successfully models the hadronic interaction at high
energies and the primary composition is dominated by heavy nuclei. Within a
standard deviation, our data points are consistent with both AGASA and HiRes.
Because of the Southern hemisphere exposure of SUGAR, this comprises
interesting support for North-South isotropy in the arrival direction of the
highest energy cosmic rays.

\noindent$(4)\quad$ An extrapolation of our analysis to the entire SUGAR data
set would involve observations using a 3 station global trigger, with its
attendant loss of angular resolution beyond 10$^{\circ}.$ Because of the
similar procedures for energy assignment in ground arrays, the consistency of
the SUGAR spectrum with AGASA observations supports recent anisotropy
studies~\cite{Anchordoqui:2003bx} of low-multipole inhomogeneities that
combine data taken from these 2 experiments.

\begin{figure}
\begin{center}
\includegraphics[height=5.8cm]{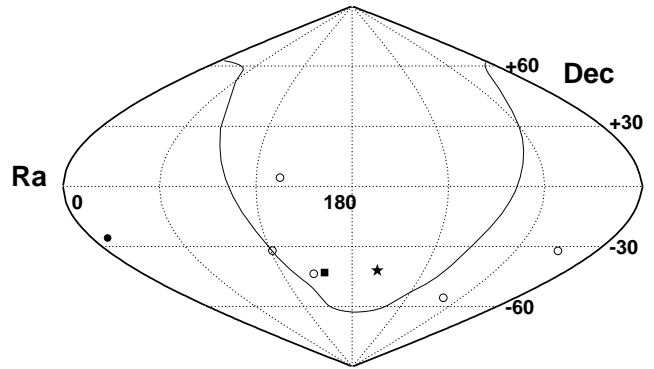}
\caption{Arrival direction of the events given in Table~\ref{tt} ($\circ$)
in equatorial coordinates B1950. Also shown in the figure are the position
in the sky of potential cosmic ray sources: Cen
A $(\star)$~\cite{Farrar:2000nw}, 
NGC3256 ({\tiny $\blacksquare$})~\cite{Giller}, and NGC253
($\bullet$)~\cite{Anchordoqui:2002dj}. The solid line indicates the
position of the Galactic plane.} \label{skymap}
\end{center}
\end{figure}

\noindent$(5)\quad$ To give a visual impression of the Southern cosmic ray
sky, in Fig.~\ref{skymap}  we show the arrival direction of the events given
in Table~\ref{tt} together with the potential sources included in the first
budget of the anisotropy search program for the Pierre Auger
Observatory~\cite{Clay:2003pv}.

\noindent$(6)\quad$ As part of our study, we have noted an interesting
differentiating signature between the {\sc qgsjet} and {\sc sibyll} models. 
Previous 
work~\cite{Anchordoqui:1998nq} has shown that for a fixed (proton) primary 
energy, muon densities obtained
in shower simulations using {\sc sibyll} fall more rapidly with lateral 
distance to the shower core
than those obtained using {\sc qgsjet}. This can be understood as a 
manifestation of the enhanced leading
particle effect in {\sc sibyll}, which can be traced to the relative 
hardness of
the electromagnetic form factor profile function. From 
Figs.~(\ref{sugarqgsjet}) and (\ref{sugarsibyll}), it is 
apparent that even
when the energy is left as a free parameter, the curvature of the distribution
$(d^2\rho_{\mu}/dr^2)$ is measurably different in the two cases, and could 
possibly serve as a discriminator between hadronic interaction models with 
sufficient statistics.

In summary, the footprints analyzed in this paper are an interesting addition 
to the data set, but not enough to weigh in on one side or the other with 
respect to the GZK question. The Pierre Auger Observatory~\cite{Sommers:jy} 
now coming on line with its  superior energy resolution and statistics will 
supply the final verdict on the GZK cutoff.

\section*{Acknowledgments} We would like to thank Tom McCauley, Tom
Paul, and Alan Watson for valuable discussions. The work of LAA
and HG has been partially supported by the US National Science
Foundation (NSF), under grants No.\ PHY--0140407 and No.\
PHY--0244507, respectively.

\end{document}